# Electrically reconfigurable nonvolatile transmissive metasurface in visible

Virat Tara[1, *], Romil Audhkhasi[1], Andrew Tang[1], Rui Chen[1,2], Johannes E. Fröch[1], Abhinav Kala[1], Juejun Hu[2], Arka Majumdar[1,3, *]

[1]Department of Electrical and Computer Engineering, University of Washington, Seattle, WA 98195, USA

[2]Department of Materials Science and Engineering, Massachusetts Institute of Technology, MA 02139, USA

[3]Department of Physics, University of Washington, Seattle, WA 98195, USA

*Corresponding authors: vtara@uw.edu, arka@uw.edu

**Abstract**

The synergy between metasurfaces and non-volatile phase change materials (PCMs) has created many reconfigurable photonic devices for applications in optical memory, optical computing and optical communications. But these advances have been limited to the infrared wavelengths due to the high loss of PCMs in the visible regime. Here we demonstrate a nonvolatile visible metasurface that is electrically reconfigurable using wide bandgap PCM $Sb_2S_3$. Our device supports a resonant mode at 610 nm, a wavelength largely under-explored for PCM-based metasurfaces. By incorporating only a 20 nm thick layer of $Sb_2S_3$, we experimentally demonstrate a resonance tuning range of 16 nm. Reversible switching of the metasurface is accomplished *in situ* using a carefully engineered, ultrathin doped silicon micro-heater. Our work paves the way for integrating PCMs into visible-frequency systems, particularly for human-centric applications such as augmented and virtual reality displays.

**MAIN TEXT**

**Introduction**

Metasurfaces utilizing subwavelength scatterers can precisely manipulate the wavefront of light, enabling them to perform complex optical functions in areas such as imaging[1][2], sensing[3], and computation[4]. However, a significant limitation of these metasurfaces is their fixed optical functionality once fabricated. Researchers explored various methods, including thermo-optic effect [5], electro-optic effect [6], liquid crystals (LCs)[7] and phase change materials (PCMs) [8][9][10], to modify metasurface's response post fabrication. All these methods essentially rely on tuning the electromagnetic (EM) modes supported by the metasurface by changing the refractive index of one of its constituent materials.

Among these above-mentioned tuning mechanisms, PCMs are particularly promising due to their non-volatile operation, which obviates the need for a constant power supply to hold the changed state. Moreover, they can provide a significant change in refractive index ($\Delta n \sim 1$) as they switch between their amorphous (a) and crystalline (c) states, as compared to other methods (for example, $\Delta n$ for LC is ~0.2[11]). This implies that they can perform the same operation in a smaller form factor without any thermal or electric-field crosstalk in steady state. Furthermore, PCMs' device-level integration can be further enhanced to provide deterministic multi-level operation by using meticulously designed heaters[12] beyond binary operations. Despite their many advantages, PCMs have been primarily limited to the near-infrared[9] and longer wavelengths[13] due to their high optical absorption in the visible spectrum[14]. Incorporating PCMs in the visible spectrum would unlock several new possibilities for applications like low-power dynamic displays[15], inkless e-papers[16] and local dimming of pixels in AR displays using unpolarized light[14]. Where, conventionally polarized dimming is used by employing liquid crystals, which leads to significant loss of light even in the transmitting state, due to the presence of polarizers.

Fortunately, recently rediscovered PCM $Sb_2S_3$ [16][17][18] is known to have a wide bandgap of 2.0 eV in its amorphous phase, making it a low-loss material for wavelengths greater than 600 nm. This, along with its significant index contrast ($\Delta n = 0.68$ at 600 nm), makes $Sb_2S_3$ highly attractive for tunable meta-optic applications in the visible. The index contrast is induced by the micro-structural phase transition of the material from a- to c- state upon heating. Past works using $Sb_2S_3$ have been primarily restricted to infrared (IR) frequencies [9][19] because of the limitations of heater materials in the visible spectrum. Some recent works with $Sb_2S_3$ in the visible [18][20] have relied on heating by a laser for inducing a phase transition in the material. While laser switching is fast, it requires bulky, precisely aligned setups that are impractical for large-scale commercial applications. Electrical control, on the other hand, provides an *in-situ* solution to the above problems.

In this work, we design and experimentally demonstrate a reversibly tunable transmissive metasurface in the visible frequency regime using a wide-bandgap PCM $Sb_2S_3$. In-situ electrical switching of the metasurface's optical response is achieved via a 55 nm thick doped microheater fabricated in a Silicon (Si) on Sapphire (SOS) substrate. By engineering the EM mode supported by the metasurface to have a strong field overlap with its PCM layer, we are able to achieve a 19 nm spectral shift in the transmission dip with only 20 nm of unpatterned $Sb_2S_3$ while minimizing the loss due to the Si heaters. Using Rapid Thermal Annealing (RTA), we are able to observe a 14 nm spectral shift in experiment accompanied by a 38% change in transmission amplitude at a wavelength of 671 nm. Finally, we perform in-situ switching of the metasurface using electrical pulses and observe a reversible change in its resonance wavelength as large as 16 nm. Our work paves the way for compact, non-volatile reconfigurable meta-optics in the visible for applications ranging from quantum photonics[21][22] to holographic displays[23][24].

**Results**
Our proposed transmissive metasurface design consists of an ultra-thin 55 nm thick layer of crystalline-Si on a Sapphire substrate followed by a 20 nm thick $Sb_2S_3$, layer, a 40 nm $Al_2O_3$ passivation layer [25] and a 100 nm metasurface made of non-stoichiometric silicon nitride $SiN_x$. Fig. 1**(A)** and 1**(B)** show the perspective and side views of the metasurface. We note that although Si is lossy in the visible wavelength range, it offers a significant advantage[2] of CMOS compatibility and ease of device-level integration over its alternatives in the visible spectrum, like Indium-Tin Oxide (ITO)[26], Graphene[27], and Silicon Carbide (SiC). A comparison of losses from Si of different thicknesses is shown in Fig. S1, indicating a relatively low absorption of 1.5% at 600 nm induced by the 55 nm thick crystalline silicon. The use of unpatterned $Sb_2S_3$ in the metasurface allows us to achieve a larger resonance shift compared to conformal $Sb_2S_3$ (see Fig. S2). The refractive index of $Sb_2S_3$ measured using ellipsometry is shown in Fig. 1**(C)** (for measurement details see Supplementary Text).

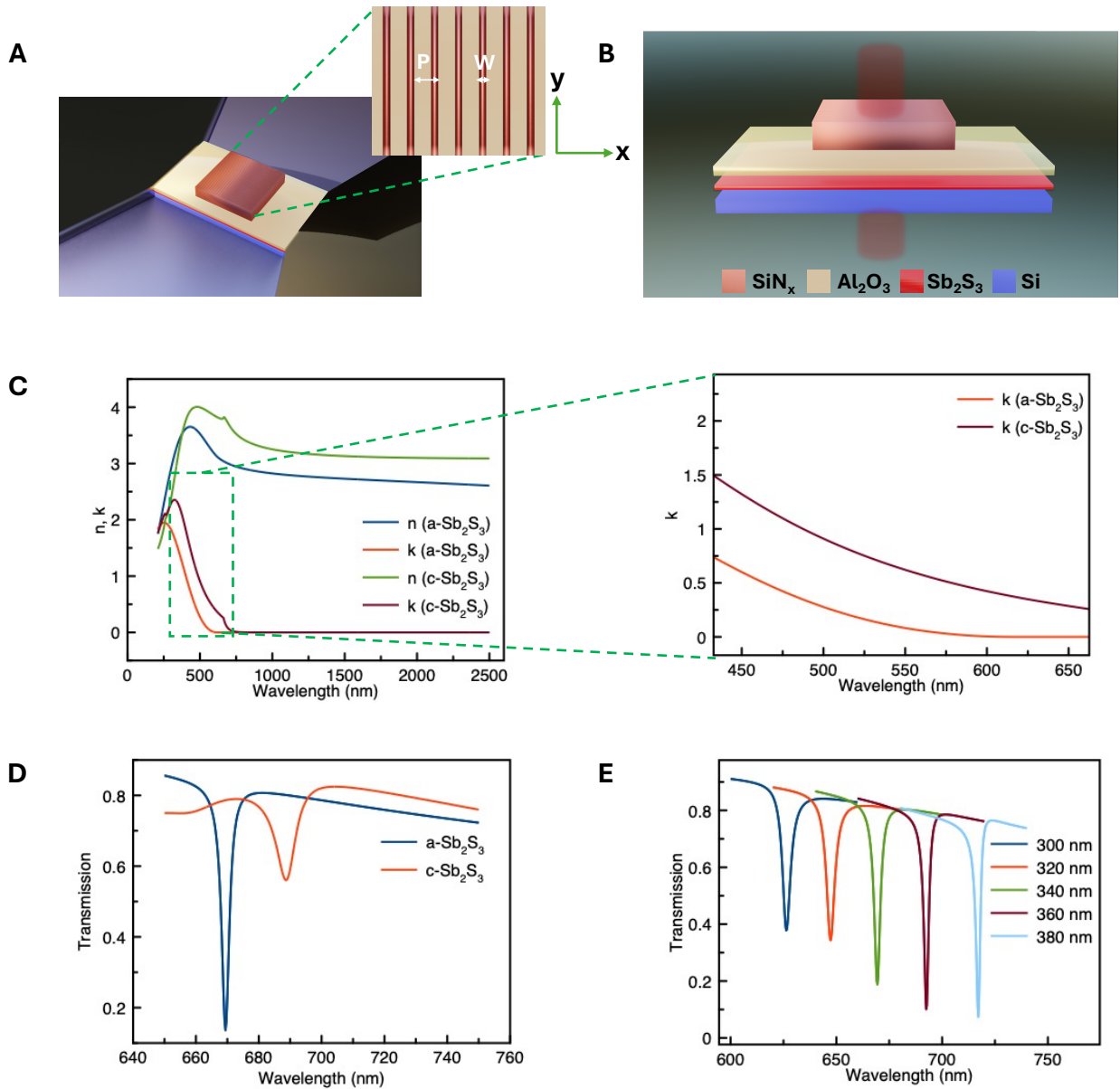

**Figure 1: Design of the reconfigurable metasurface.** **(A)** Perspective view of the device with the metasurface and the electrodes. Zoomed-in is the top view of the $SiN_x$ periodic grating with width (W) and period (P). **(B)** Side view of the metasurface with its 4 different layers. **(C)** Measured refractive index of $Sb_2S_3$. The zoomed in plot on the right displays the losses of $Sb_2S_3$ in visible wavelengths. **(D)** Simulated transmission spectra when $Sb_2S_3$ is in its a- and c- states. **(E)** Transmission spectrum in simulation as we sweep the period of the $SiN_x$ grating.

The unit cell, with periodic boundary conditions applied to the in-plane directions and a Perfectly Matched Layer (PML) in the out-of-plane direction, was simulated using Lumerical FDTD software. The device was illuminated using a plane wave source. The incident electric field polarization is maintained along the x-axis as shown in Fig. 1**(A)**. The refractive index of Si and Sapphire were used from the built-in Lumerical material library. Refractive index for the capping layer (alumina) and $SiN_x$ (for films deposited in-house) were measured using an ellipsometer. The values for the real part of refractive index (n) were roughly 1.63 and 2.0 for Alumina and $SiN_x$. Fig. 1**(D)** presents the spectrum of the meta-optics with a periodicity (P) of 340 nm and the grating width (W) of 100 nm. The simulated resonance wavelength redshifts by 19 nm as we switch the refractive index of $Sb_2S_3$ from a- to c- state. This resonance shift is accompanied by a decrease in the quality factor (Q-factor) from 230 to 102, due to the higher loss of $Sb_2S_3$ in the c- state. The resonant mode when in a- state also supports a $2\pi$ phase shift in reflection[28], as shown in Fig. S3. Therefore, this structure can also be used for phase modulation applications in reflection[29]. Fig. 1**(E)** presents the change in resonant wavelength as the periodicity of the meta-atom is swept from 300 to 380 nm. This linear scaling of the resonance wavelength to periodicity helps in setting the desirable resonant wavelength during fabrication.

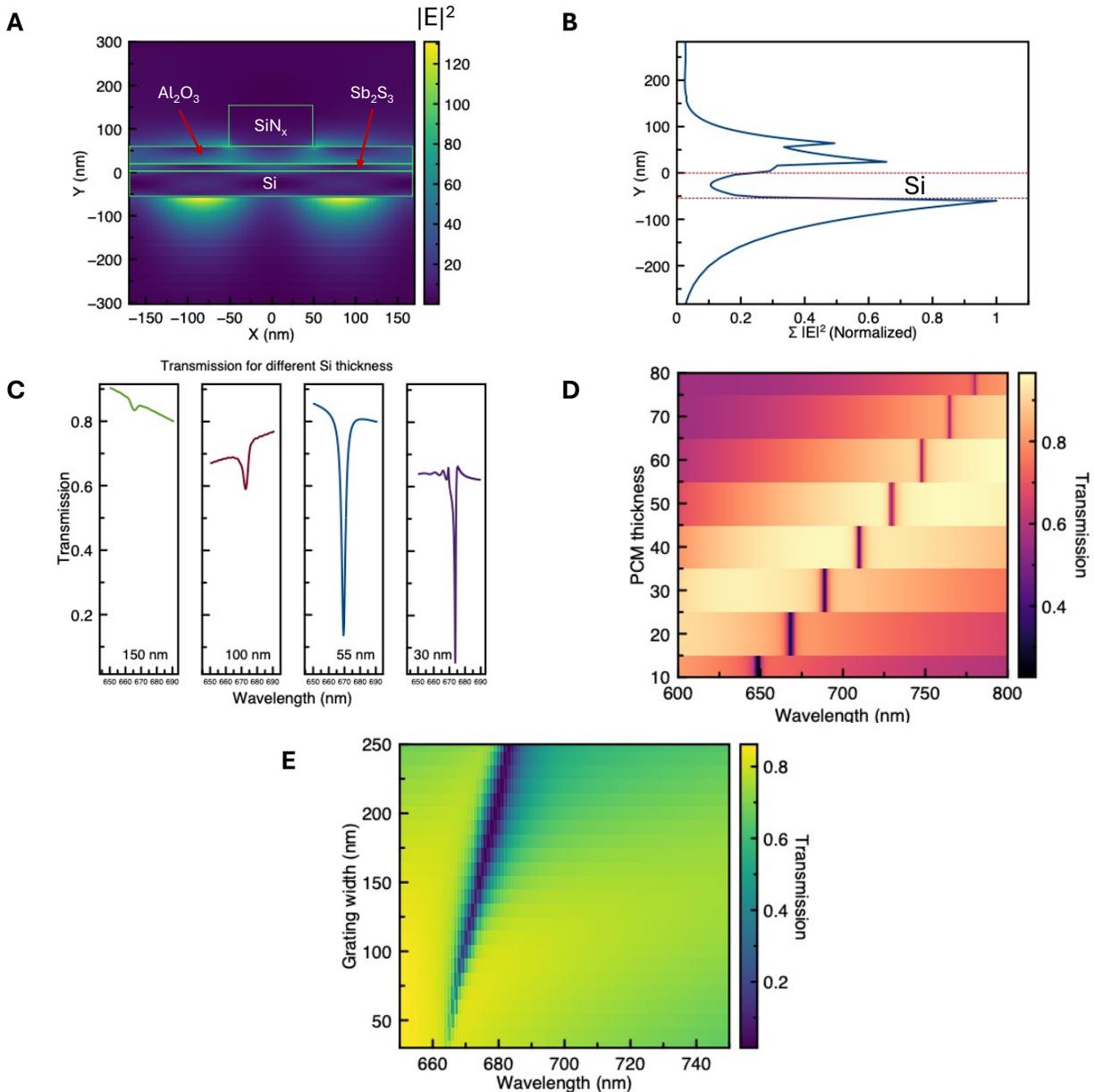

**Figure 2: Design optimization. (A)** The $|E|^2$ inside the unit cell at resonance. **(B)** Sum of the E-field intensity along the thickness of the device. The red dotted lines depict the Si layer. **(C)** The spectra for different values of Si slab thickness. **(D)** The change in Transmission spectra of the device as the thickness of the PCM layer is varied. **(E)** Simulated Transmission spectra for different grating widths.

Further simulations were performed to analyze and optimize the metasurface response. Fig. 2**(A)** shows the field profile of the Transverse Magnetic (TM) mode inside the cavity at the resonant wavelength when $Sb_2S_3$ is in a- state. The incident electric field is polarized along the x-direction. Due to the ultra-low thickness of the Si slab the mode is loosely confined in the Si slab and therefore it largely resides outside the high-index, lossy Si slab. Furthermore, due to the EM boundary conditions we see a step increase in field inside the $Sb_2S_3$ layer given by the following relation $E_{SbS} = \left(\frac{n_{Si}}{n_{SbS}}\right)^2 \times E_{Si}$. In contrast, increasing the Si thickness increases the mode overlap with the lossy Si slab[30]. Fig. 2**(B)** presents the field intensity summed along the *x* axis of the device, showing minimal electric field inside the Si layer.

Fig. 2**(C)** presents the simulated transmission spectrum for four different Si layer thicknesses. To ensure that the transmission dip shows up at approximately a wavelength of 670 nm, we tuned the grating period from 380 to 200 nm for Si layer thicknesses varying from 30 to 150 nm while keeping the other parameters fixed. We observe that as the Si layer thickness increases, the transmission dip becomes broader. This can be attributed to an increase in the resonant mode overlap with the Si layer for larger thicknesses (see Fig. S4 for the electric field intensity profile in a metasurface with a Si layer thickness of 150 nm). The Q-factors for the 150 nm, 100 nm, 55 nm, 30 nm layer thickness are 116, 182, 230 and 817 respectively. We note that while a 30 nm Si layer gives the highest Q-factor of 817, we used 55 nm thick Si to avoid an increase in resistance with decreasing Si thickness. We show in Fig. S5 that a 30 nm thick Si layer can be integrated to make a metasurface guided mode resonator (GMR) operating over the whole visible wavelength range from 450 to 750 nm.

Fig. 2**(D)** presents the transmission spectra with varying thickness of the PCM layer. The resonance shifts to a longer wavelength with a thicker PCM. A 20 nm thickness was picked to produce a resonant wavelength well within the visible frequency range. This resonance also shows a high contrast of 72% at 670 nm wavelength, where the contrast is calculated as $(T_{max} - T_{min})/(T_{max} + T_{min})$. We emphasize that the resonance allows such a thin layer of PCM to produce a large resonance shift. A thin PCM layer is advantageous for reversible switching, as a larger volume of PCM suffers more from issues such as temperature nonuniformity and thermal reflowing. Besides, the melt and quench process demand a fast enough quench rate of $\sim 10^{-9} K/s$, ultimately limiting the maximum thickness of the PCM layer[14]. Fig. 2**(E)** shows the spectrum of a-$Sb_2S_3$ with varying grating width, indicating its strong effect on both extinction ratio and linewidth. We design the grating width as 100 nm because it can support a resonance with a high Q-factor of 230, while remaining compatible with the critical dimension of our lithography tool. More details on the fabrication of the sample are given in the Methods section.

Several devices of different sizes were fabricated on the same chip. Fig. 3(A) shows the micrograph of a $100 \times 100$ μm$^2$ device before and after Rapid Thermal Annealing (RTA) at 325°C for 10 mins under an inert environment. The formation of large crystals is evident under the microscope. The different colors of $Sb_2S_3$ crystals could be attributed to anisotropic nature of $Sb_2S_3$[31]. Fig. 3**(B)** presents the Scanning Electron Microscope (SEM) image of one of the devices with a period of 350 nm, showing a smooth side wall. Devices with different periods were measured in a transmission setup, see Fig. S6 for description of the setup. A 10× objective with NA 0.26 was used to illuminate the sample. A confocal pinhole was used to further reduce the angle of the incident light cone on the sample, allowing only the desired device to be illuminated for measurement with near normal angle of incidence. The transmitted light was then collected using a 50× objective with NA 0.65 and fed to a spectrometer.

Using the setup, we first measured devices with four different periods (300 nm, 320 nm, 350nm, and 370 nm) with the PCM in a- state. As expected from simulations, an increase in the period of the unit cell results in a red shift of the resonant mode as shown in Fig. 3**(C)**. In Fig. 3**(D)**, we show the spectrum of the 100 × 100 µm² device before and after RTA. The calculated transmission contrast at 671 nm is 38% and the red shift in resonance is 14 nm. The experimental Q-factors calculated by fitting the spectrum using the Fano model[32] were found out to be 80 and 57 for $Sb_2S_3$ in a- and c- state respectively. The lower Q factors in experiment as compared to simulation could be because of additional losses of doped Si as compared to undoped Si and non-ideal incident angle due to the presence of an objective.

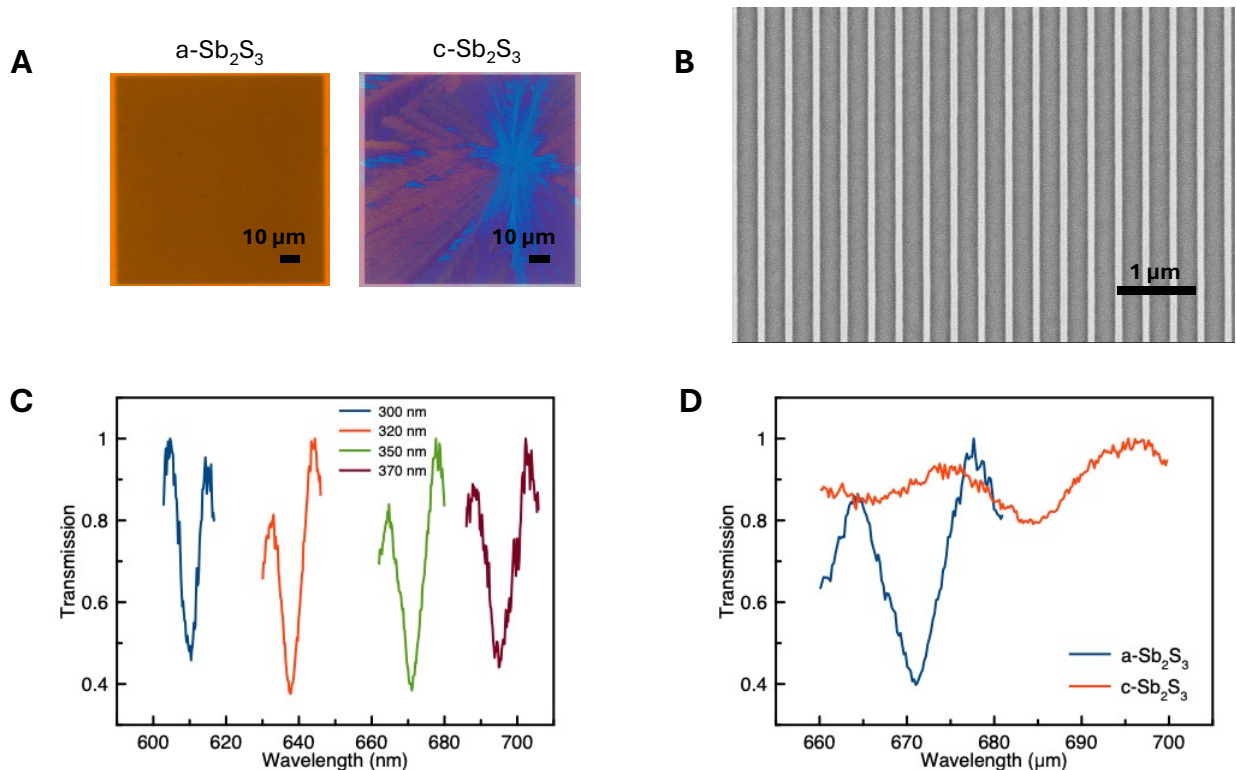

**Figure 3: Measurements of fabricated device. (A)** Micrographs of the 100x100µm² devices before (left) and after (right) RTA. **(B)** The Scanning Electron Microscope image of one of the fabricated devices. **(C)** The spectrum of different devices with increasing unit cell periodicity. **(D)** The spectrum of the device before and after RTA.

Lastly, we perform electrical switching of the metasurface through the in situ doped silicon heaters and two metal contact pads (see Methods section for the fabrication process). Electrical signals were applied by an arbitrary function generator through two probe positioners, which are integrated with the free-space setup. A picture of the setup is shown in Fig. S7. Since the power required to attain the switching temperatures scales with the area of the device, we attempted to switch devices with a smaller area first. Fig. 4(A) shows a 17 × 17 μm² device that was switched under a microscope. We observed a reversible change in color upon applying 30V 3.5μs (12.6 μJ) pulse with 21ns rise and fall times, and 15.4V 1μs (0.96 μJ) pulses with 21ns rise and fall times at 500 kHz for amorphization and crystallization, respectively, as shown in Fig. 4(B). Video recordings of the switching under microscope are given in Movie S1&2. We also show the optical micrographs of the same device under different crystallization pulse conditions in Fig. S8.

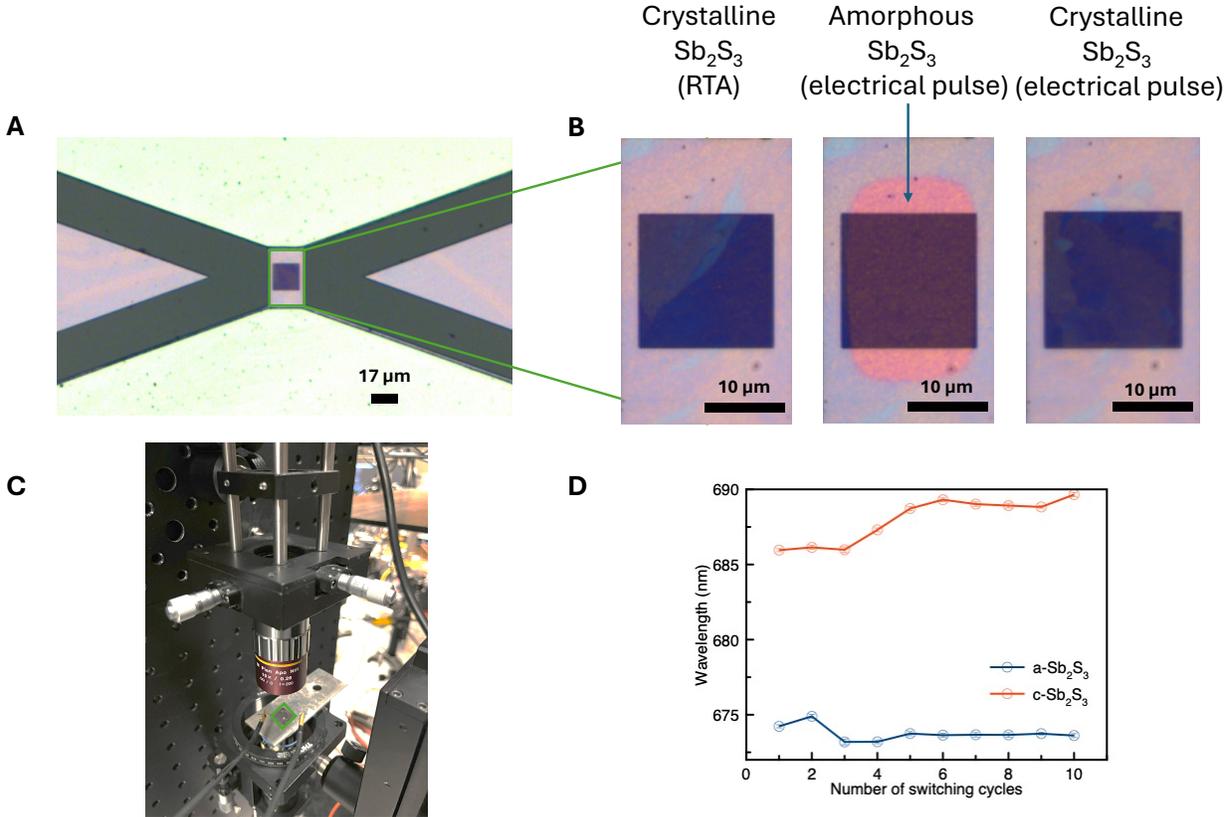

**Figure 4: Reversible switching using electrical pulses. (A)** The micrograph of a 17x17μm² device with contact pads for electrical switching. **(B)** The micrographs of the device in **(A)** after RTA (leftmost), after application of amorphization pulse (center), and after application of crystallization pulse (rightmost). **(C)** The transmission setup with probe stationers used for the measurements. The highlighted portion is the sample. **(D)** The 10 cycles of switching of $Sb_2S_3$ and the resonance shift induced by it. Error bars depict the data fitting errors.

Following this, we switched a larger (20 × 20 μm²) device with a unit cell period of 350 nm in the transmission setup shown in Fig. 4(C). The electrical pulses applied for inducing amorphization and crystallization were 25V 9μs (22.5 μJ) with a 21ns rise and fall time and 12.5V 100ms (62.5mJ) with a 21ns rise and fall time, respectively. We reversibly switch the device for 10 cycles and plot the resonance shift in Fig. 4(D). The slight change in the resonance wavelength in c- state over the cycles is attributed to multiple crystalline phases of $Sb_2S_3$[16]. Additionally, recrystallization of the material after amorphization results in slightly different grain sizes and orientations as seen in Fig. 4(B) on comparing the leftmost and rightmost micrograph. The resonance shift corresponding to the last cycle is 16 nm. We note that the device did not become dysfunctional after 10 cycles but shows significant structural changes under the microscope. The micrograph of the device after 10 cycles and transmission spectrum for 10 cycles is given in Fig. S9 and S10. The structural changes are attributed to the dewetting of the PCM at the time of amorphization, when the material is in its liquid state. This issue could be resolved by restricting the movement of the material in the in-plane direction by forming isolated PCM meta-atoms, as proposed in [33], or by using a fishnet type metasurface design which is more suitable for growth dominated PCMs[34]. Beyond 1D resonance, we also designed and measured an ultra-thin 2D GMR metasurface, with results presented in the Fig. S11.

**Discussion**
Future work include solving the issue of structural change due to dewetting, such as by patterning the PCM into smaller patches[33] or by choosing optimal capping layers[35] with proper wetting properties. As visible in Movie S2 and as reported earlier[36] the crystallization of the material is dominated by growth rather than by nucleation. Therefore, it is desired to increase the nucleation density of $Sb_2S_3$, which speeds up the crystallization process and improves the material reliability. One potential solution is to use a thin ZnS layer between the substrate and the PCM, which reduces the crystal size significantly from 10 μm to 2-3 μm [36]. As another future research direction, it is highly desired to extend the transparency window of PCMs to the green and blue wavelengths. PCMs like MnTe [37][38] with an even wider band gap (2.7eV) than $Sb_2S_3$ need to be investigated further for integration into photonic platforms. We also provide a comparison of our work to other non-volatile reconfigurable metasurfaces in Table S1.

In conclusion, we demonstrated visible-frequency reconfigurable metasurfaces using a wide bandgap PCM $Sb_2S_3$. An ultra-thin layer of 20 nm PCM provides a resonance shift of as large as 16 nm. The metasurface is tunable with carefully engineered, thin doped Si microheaters in the visible frequency regime, despite its inherent losses. We then validated the functionality our design through measurements of the fabricated devices, demonstrating resonance spectra shift with different unit cell periodicity. The resonance shift of around 16 nm was verified with RTA and followed by in situ reversible switching of the PCM through the integrated heaters, demonstrating over 10 switching cycles. Our work paves the way for applications of PCMs in the visible frequency regime and can enable development of next-generation displays.

**Materials and Methods**
Fabrication: The devices were fabricated on a 230 nm-doped silicon-on-sapphire wafer. Firstly, Si was doped using Phosphorus ions and annealed, resulting in a uniform doping concentration of $10^{20}$cm$^{-3}$ along the depth of Si. Then, the Si layer was thinned down to 55 nm using reactive ion etching (RIE) with fluorine gas (Oxford PlasmaLab 100). Then, 20 nm of $Sb_2S_3$ (sputtering target from PlasmaMaterials) was deposited using sputtering (Lesker Lab18) and covered with 40 nm $Al_2O_3$ using atomic layer deposition (Oxford PlasmaLab 80plus). At the very top, 100 nm of $SiN_x$ was deposited using plasma-enhanced chemical vapor deposition (SPTS SPM). Finally, the top $SiN_x$ layer was patterned using 100kV electron-beam lithography (JEOL 6300FS) and etched using RIE with Fluorine gas (Oxford PlasmaLab 100). To put metal pads for electrical switching, the $SiN_x$, $Sb_2S_3$, and $Al_2O_3$ layers were removed from the area to form electrical contact between metal and doped silicon. NR9G-3000PY photoresist was used as an etching mask and exposed using laser direct write lithography (Heidelberg DWL66+). The films were removed using RIE with Fluorine gas for SiN$_x$ and Chlorine gas for Sb$_2$S$_3$ and Al$_2$O$_3$. Finally, Ti/Pt were deposited using sputtering (Evatec LLS EVO) and lifted off using NR9G-3000PY again, patterned using laser direct write lithography. The final device resistance was ∼250 Ohms for both 17 × 17 and 20 × 20 µm² devices.

Setup: The sample was illuminated using a broadband laser (Fianium WhiteLase Micro). The laser light was first collimated using Lens 1 and then focused onto the sample using a 10x objective (Objective 1) with a numerical aperture (NA) of 0.26. The Dichroic Mirror 1 is used to deflect the reflected light from the metasurface to a camera along with an imaging lens (Lens 2). The transmitted light from the sample is collected using a 50× objective (Objective 2) with NA 0.65 and again split into another camera with an imaging lens (Lens 3). Finally, the laser light is fiber coupled using Lens 4 and fed to the spectrometer (IsoPlane SCT 320 by Princeton Instruments). The electrical measurements were carried out using a pair of electrical probes on two probe positioners (Cascade Microtech DPP105-M-AI-S). The electrical pulses were generated by a pulse function arbitrary generator (Keysight 81160A) and amplified using a Cleverscope CS1070 with a 50 Ω output connected to the 250 Ω load (our device). The voltage drop on the device was calculated as $V_{device} = V_{applied} \times \frac{R_{device}}{R_{device}+R_{out}}$.

**Acknowledgements**

**Funding:** The work is supported by DARPA-ATOM program. Part of this work was conducted at the Washington Nanofabrication Facility / Molecular Analysis Facility, a National Nanotechnology Coordinated Infrastructure (NNCI) site at the University of Washington with partial support from the National Science Foundation via awards NNCI-1542101 and NNCI-2025489.

**Author contributions:** AM, RC and VT conceptualized the project. VT performed the simulations, fabrication and experiment. RA helped with the simulations. AT helped with the fabrication. Setup used for measurements was built by JF and AK. AM and JJ supervised the project. Models used to fit ellipsometry data were made by RC. VT wrote the manuscript with input from all the authors.

**Competing interests:**
Authors declare that they have no competing interests.

**Data and materials availability:**
All data are available in the main text or the supplementary materials.

Supplementary Materials for

**Electrically reconfigurable nonvolatile transmissive metasurface in the visible**

Virat Tara[1,*], Romil Audhkhasi[1], Andrew Tang[1], Rui Chen[1,2], Johannes Froech[1], Abhinav Kala[1], Juejun Hu[2], Arka Majumdar[1,3,*]

[1]Department of Electrical and Computer Engineering, University of Washington, Seattle, WA 98195, USA
[2]Department of Materials Science and Engineering, Massachusetts Institute of Technology, MA 02139, USA
[3]Department of Physics, University of Washington, Seattle, WA 98195, USA

*Corresponding authors: vtara@uw.edu, arka@uw.edu

**This PDF file includes:**

    Supplementary Text
    Figs. S1 to S11
    Tables S1
    Movies S1 to S2

**Other Supplementary Materials for this manuscript include the following:**

    Movies S1 to S2

**Supplementary Text**

Refractive index of $Sb_2S_3$

The refractive index was measured on a Woollam RC2 ellipsometer on a 20 nm thick film deposited via sputtering on a Silicon substrate. The measured wavelength range was 210 to 2500nm. Measurements were taken at 55º, 65º and 75º angle of incidence. One Cody-Lorentz and one Lorentz model were used to perform the fitting for a-$Sb_2S_3$. Two Cody-Lorentz models were used to perform the fitting for c-$Sb_2S_3$. Mean squared error in both the cases were 1 and 6 respectively. A 40 nm thick $SiO_2$ cap layer was also accounted for in the ellipsometry model in the case of c-$Sb_2S_3$.

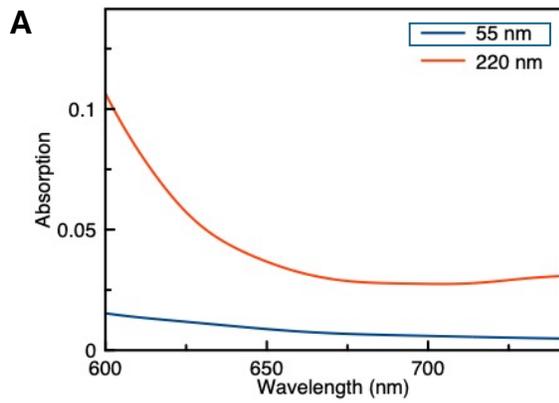 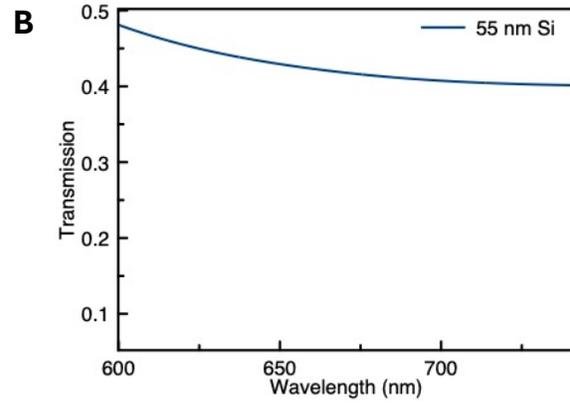

**Fig. S1. Absorption and transmission of 55 nm Silicon on Sapphire.** (**A**) Simulated absorption of different thicknesses of Silicon on Sapphire. Absorption was calculated as 1 - (Reflection + Transmission). (**B**) Simulated transmission of 55 nm thick Silicon on Sapphire.

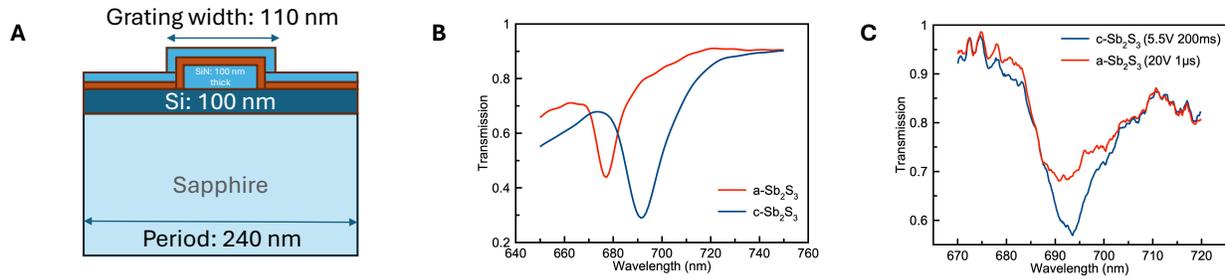

**Fig. S2. Device with conformal Sb₂S₃. (A)** Device schematic with conformal 20 nm thick Sb$_2$S$_3$ capped with 40 nm Al$_2$O$_3$. **(B)** Change in resonance in simulation $\Delta\lambda$ = 12 nm. **(C)** Change in resonance in experiment $\Delta\lambda$ = 3nm. We saw a lower-than-expected resonance change in the experiment as compared to the simulation. This could be attributed to temperature gradients along the thickness of the Silicon Nitride grating, which prevents the material from transitioning to the completely amorphous phase. Furthermore, increasing the amorphization pulse condition leads the device to get damaged.

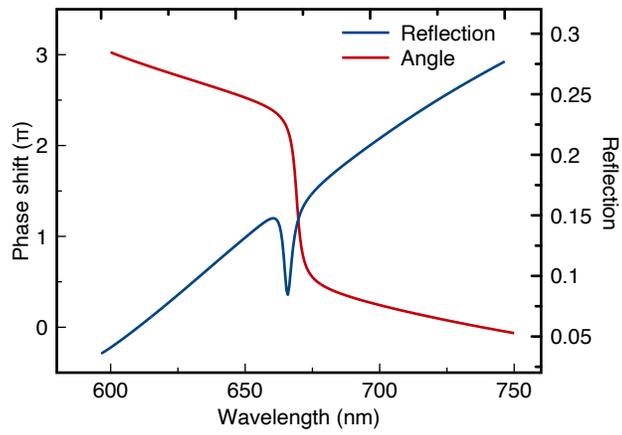

**Fig. S3. 2π phase shift at resonance in the proposed metasurface.** The simulated phase shift and reflection spectrum at resonance in the proposed metasurface. Despite being thin, our proposed metasurface supports a 2π phase shift in reflection.

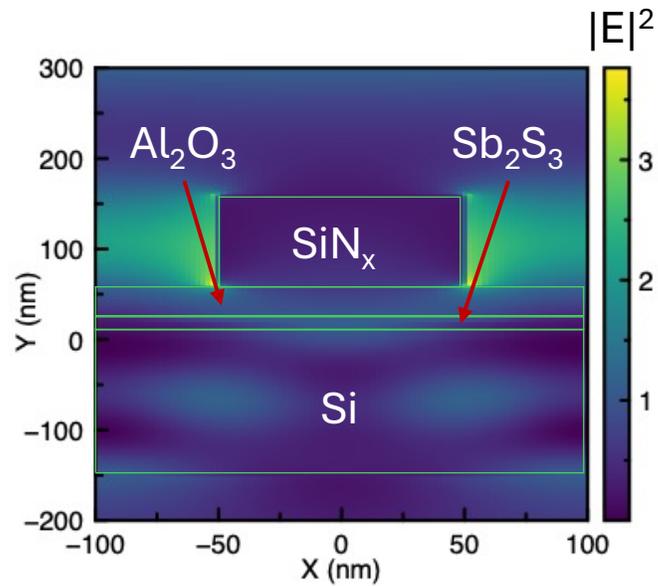

**Fig. S4. Electric field inside a 150 nm thick Si resonant metasurface.** Simulated electric field inside a metasurface with 150 nm thick Si. Due to the large thickness of Si, the metasurface is hardly able to support a resonant optical mode, as also shown in Fig. 2**(B)**.

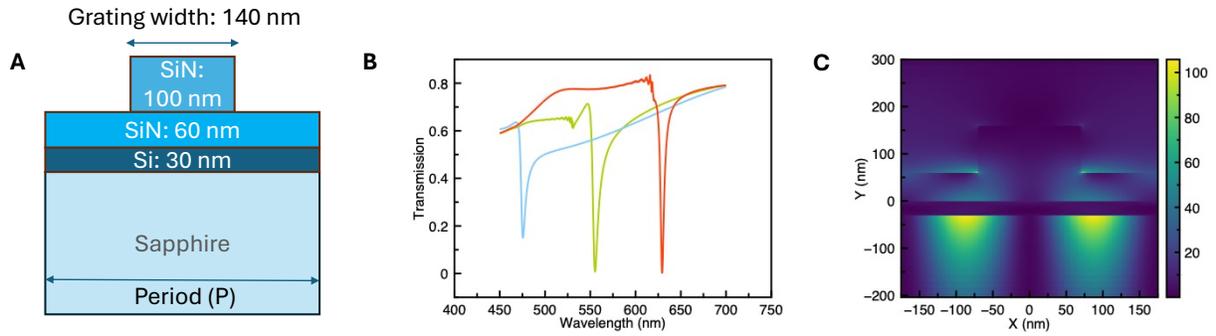

**Fig. S5. 30 nm Si based guided mode resonance (GMR) design for the full visible spectrum. (A)** Side view of the unit cell schematic of the GMR. **(B)** Simulated transmission spectrum for periods 240 nm, 300 nm and 350 nm. **(C)** Shows the $|E|^2$ of the mode of a device with period 350 nm.

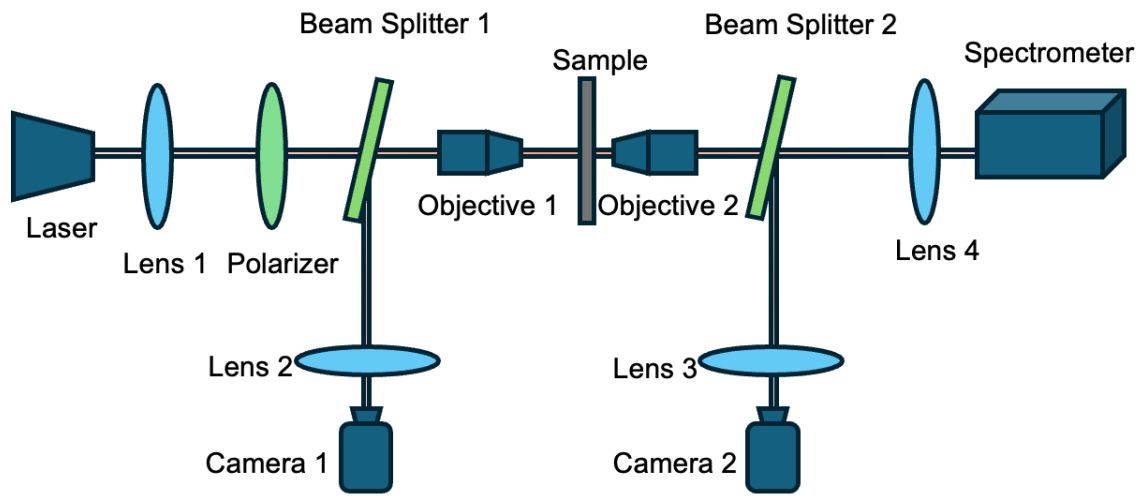

**Fig. S6. Schematic of the measurement setup used for transmission measurements.**

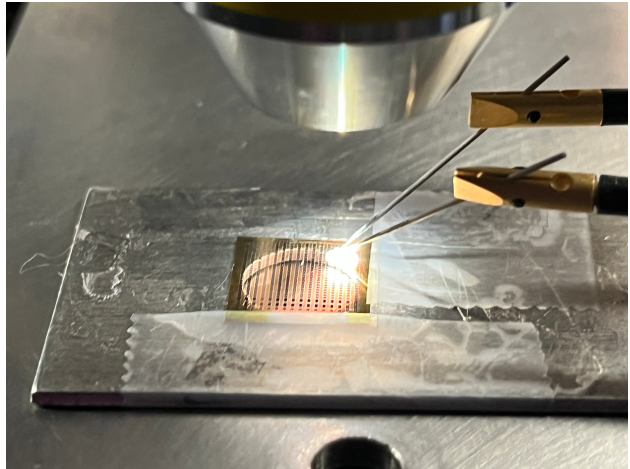

**Fig. S7. Setup for under microscope switching of Sb$_2$S$_3$ for Fig. 4 (B).**

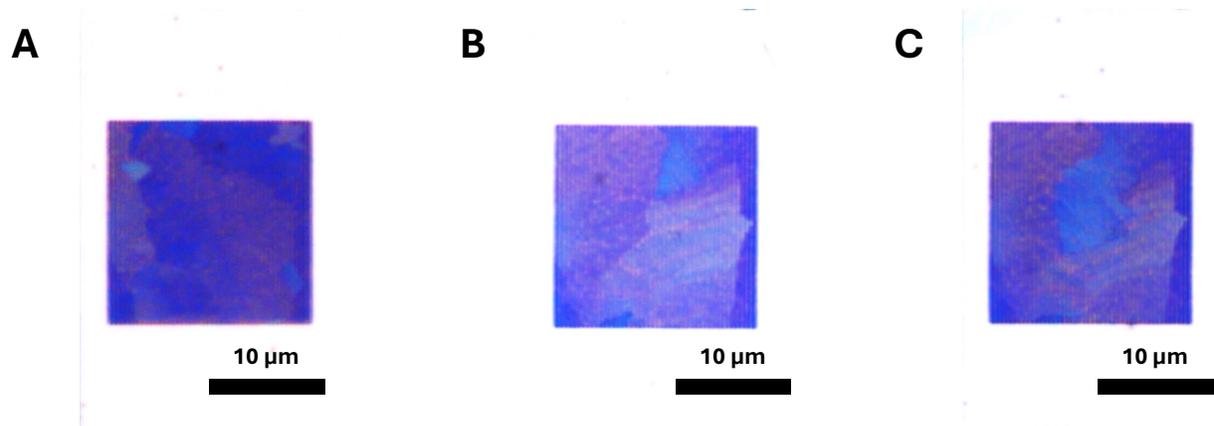

**Fig. S8. Device micrograph after different crystallization pulses. (A)** Micrograph of a device that was switched by applying a periodic pulse of frequency 500 kHZ and duty cycle 50%. **(B)** and **(C)** Micrograph of the same device after it was crystallized by applying a 100ms pulse with 21 ns rise and fall time at 2 different switching cycles. The images were enhanced to clearly show differences in the $Sb_2S_3$ layer upon crystallization with the given pulse conditions pulse conditions. We note that periodic pulses results in a smaller grain size as compared to single pulses of 100 ms width.

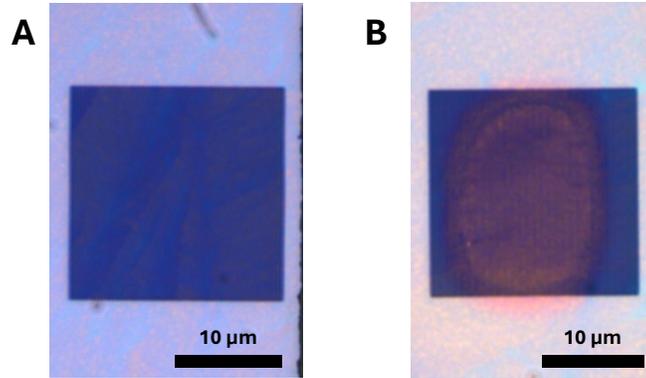

**Fig. S9. Device micrograph after 10 cycles. (A)** Micrograph of a device that hasn't been electrically switched yet. **(B)** The micrograph the device used for Fig. 4**(D)** after 10 switching cycles.

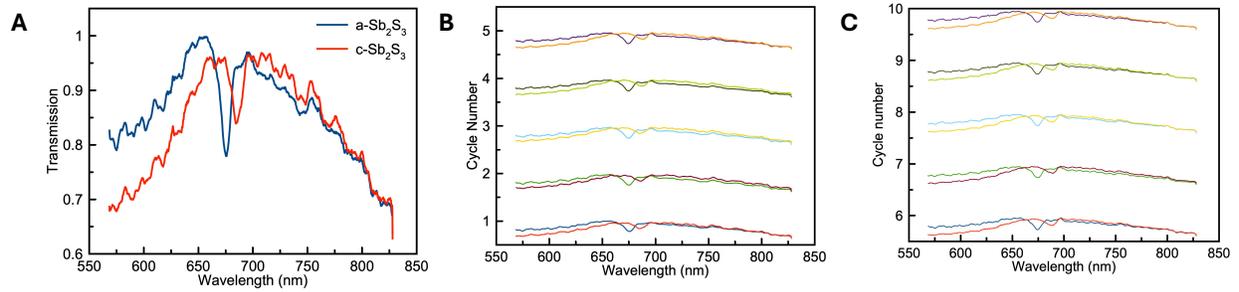

**Fig. S10. Transmission spectrum for reversible switching cycles. (A)** The transmission spectrum for one switching event. **(B)** & **(C)** Show the transmission spectrum for 10 cycles. We note that the transmission at resonance for this device is higher as compared to the device used in Fig. 3 **(D)**. This is because the device used for Fig. 3 **(D)** measurement is 25 times larger than the one used here. Since GMRs are non-local optical modes, generally a larger device is preferable since that gives a result closer to the simulation.

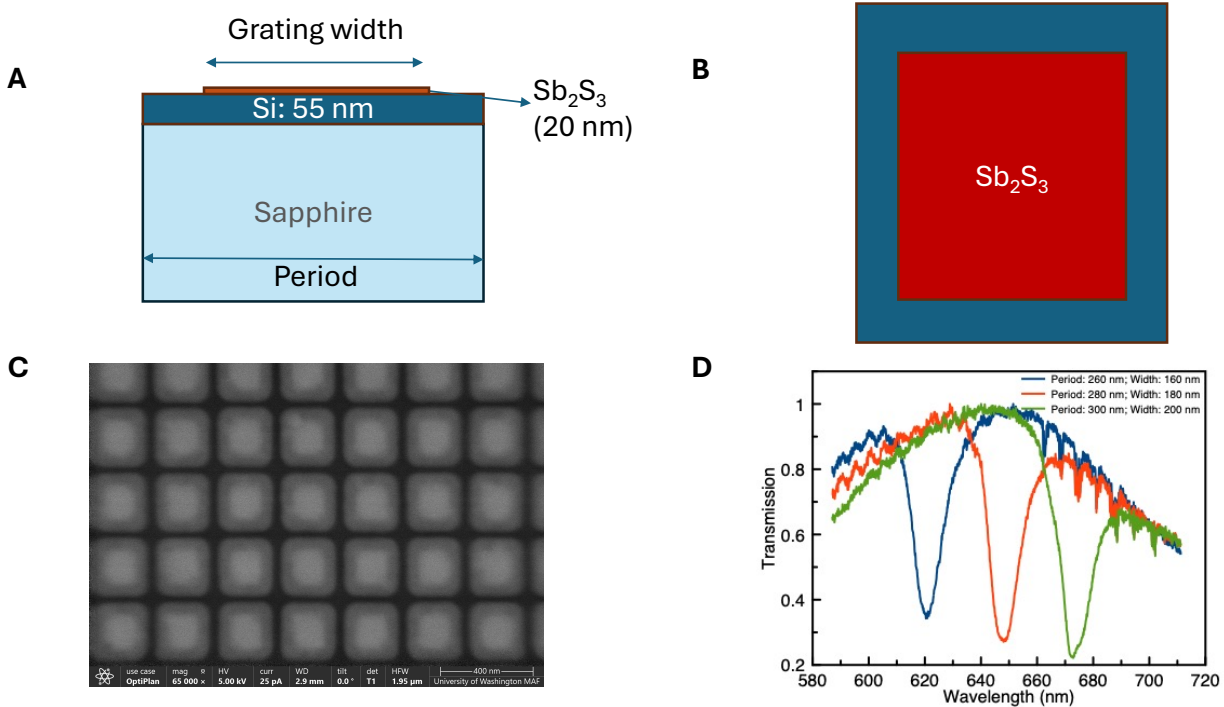

**Fig. S11. 2D Guided Mode Resonance (GMR) design and demonstration. (A)** The side view of the schematic for a 2D GMR. **(B)** Top view of the proposed 2D GMR device. **(C)** Scanning electron microscope (SEM) of the fabricated metasurface. A conformal 40 nm $Al_2O_3$ was used to protect the PCM. **(D)** Transmission measurement of 3 different devices with different periodicities. Since the crystallization of $Sb_2S_3$ appears to be growth dominated instead of being nucleation dominated. We found it be very difficult to crystallize all the nano-patches and hence reversibly switch this metasurface.

**Table S1. Comparison of our work to other non-volatile reconfigurable metasurfaces.**

| Reference link | Phase Change Material (PCM) | Reconfigurable device area | PCM switching conditions | Switching cycles | PCM thickness | Electrical/ Optical switching | Wavelength range |
|---|---|---|---|---|---|---|---|
| [1] | GST (225) | 10.5 x 10.5 µm² | Set: 1.7V 200µs<br><br>Reset: 3.8V 200ns | 50 | 40 nm | Electrical (W heater) | NIR (1370 nm - 1640 nm) |
| [2] | GST (225) | NA for optical switching | Set: 10.5µJ<br><br>Reset: 0.28nJ | 8 | 25 nm | Optical | Visible, MWIR, LWIR |
| [3] | GST (225) | NA for optical switching | Set: 60mJ cm²<br><br>Reset: 160mJ cm² 100ns pulse | 10 | 70 nm | Optical | 3-5 µm |
| [4] | GST (225) | NA | Set: 1.25nJ<br><br>Reset: 0.39nJ | - | 70 nm | Optical | 730 nm |
| [5] | $Ge_2Sb_2Se_4Te$ | 140 x 140 µm² | Set: 7.7J<br><br>Reset: 0.48mJ | 1250 | 370 nm | Electrical | 1430 nm |
| [6] | $Ge_2Sb_2Se_4Te$ | 200 x 200 µm² | Set: 2.5J<br><br>Reset: 100µJ | 40 | 250 nm | Electrical | 1550 nm |
| [7] | $Sb_2Se_3$ | 30 x 30 µm² | Set: 11.1µJ<br><br>Reset: 1.9µJ | 1000 | 20 nm | Electrical | 1518 nm |
| [8] | $Sb_2S_3$ | 15 x 15 µm² | Set: >1J<br><br>Reset: 8µJ | 9 | 20 nm | Electrical | ~ 1150 nm |
| [9] | $In_3SbTe_2$ | NA | Set: 3.6µJ<br><br>Reset: 2.8nJ | 20 | 50 nm | Optical | 4.66 µm |
| [10] | GST (225) | 17 x 17 µm² | Set: 9.4µJ (Slow) 0.46µJ (Fast)<br><br>Reset: 0.68µJ | 26 | 10 nm | Electrical | 3-5 µm |
| This work | $Sb_2S_3$ | 20 x 20 µm² | Set: 62.5mJ<br><br>Reset: 22.5µJ | 10 | 20 nm | Electrical | Visible |

**Movie S1.**

Change in the device under a microscope on applying an amorphization pulse (30V 3.5µs pulse with 21ns rise and fall times).

**Movie S2.**

Change in the device under a microscope on applying a crystallization pulse (15.4V 1µs pulses with 21ns rise and fall times at 500 kHz).